\documentclass[letterpaper,journal]{IEEEtran}

\usepackage{amsmath,amsfonts}
\usepackage{array}
\usepackage{textcomp}
\usepackage{stfloats}
\usepackage{url}
\usepackage{color}
\usepackage{verbatim}
\usepackage{graphicx}
\usepackage[absolute,overlay]{textpos}
\usepackage{ragged2e}
\usepackage{balance}
\usepackage{hyperref}
\usepackage{multirow}
\usepackage{makecell}
\usepackage{amsthm}
\usepackage{booktabs}

\usepackage{algorithm}
\usepackage{algorithmic}

\usepackage{tabularx}

\usepackage{caption}
\usepackage{subcaption}

\begin{document}
	
	\title{Intelligent Multimodal Multi-Sensor Fusion-Based UAV Identification, Localization, and Countermeasures for Safeguarding Low-Altitude Economy}
	
	\author{Yi Tao, Zhen Gao, Fangquan Ye, Jingbo Xu, Tao Song, Weidong Li, Yu Su, Lu Peng, Xiaomei Wu, Tong Qin, Xiao Liang, Zhongxiang Li, Dezhi Zheng}
	
	\maketitle
	
	\begin{abstract}
		With the growing prominence of UAV safety issues in the low-altitude economy, accurate identification, localization, and countermeasures have become core priorities for airspace security. In response, this paper introduces a deep learning-based solution that integrates multimodal sensor fusion to collaboratively execute these critical functions.
		By incorporating deep learning technologies, the UAV management system synthesizes radio frequency feature analysis, radar detection, opto-electronic sensing, and other approaches to achieve the identification of UAVs.
		For localization, the system leverages multi-sensor data fusion and artificial intelligence (AI) technologies, integrated with an air-space-ground integrated communication network, to perform real-time tracking of UAV flight dynamics, providing robust support for early warning and decision-making.
		At the countermeasure level, it adopts comprehensive measures integrating ``soft kill'' and ``hard kill'', encompassing electromagnetic signal jamming and physical interception. By integrating AI technologies, it aims to achieve both the efficiency improvement of individual countermeasure approaches and dynamic, multi-modal decision-making based on multiple countermeasures, which significantly enhances the response efficiency and disposal accuracy of low-altitude UAV management.
	\end{abstract}
	

	\section{Introduction}
	With the rapid implementation of application scenarios such as logistics and delivery, power inspection, and urban air mobility, the low-altitude economy (LAE), an emerging industry integrating airspace resources with economies, is entering an explosive phase of large-scale development \cite{1,11015739,liu2024near}.
	As shown in Figure 1, the integration of communication and sensing capabilities in low-altitude uncrewed aerial vehicles (UAVs) enables flexible adaptation to diverse scenarios, demonstrating their critical role and value in the LAE \cite{fan2020review,yu2024performance}.
	Its potential to generate significant economic and social benefits has garnered widespread anticipation \cite{9892691}.
	However, severe new safety challenges lie behind the booming development of the LAE.
	For instance, the chaos of UAVs remains rampant. Incidents of unregistered UAVs intruding into sensitive areas such as airport clearance zones are common occurrences, posing a serious threat to public safety \cite{10759668}.
	Additionally, the demand for airspace usage has led to a rising risk of conflicts between UAVs, while traditional airspace management systems struggle to cope with the complex and dynamic air traffic situation \cite{10266978}.
	Such core risks have become critical bottlenecks restricting the development of the LAE.
	
	\begin{figure}[htbp]
		\centering
		\includegraphics[width=1\linewidth]{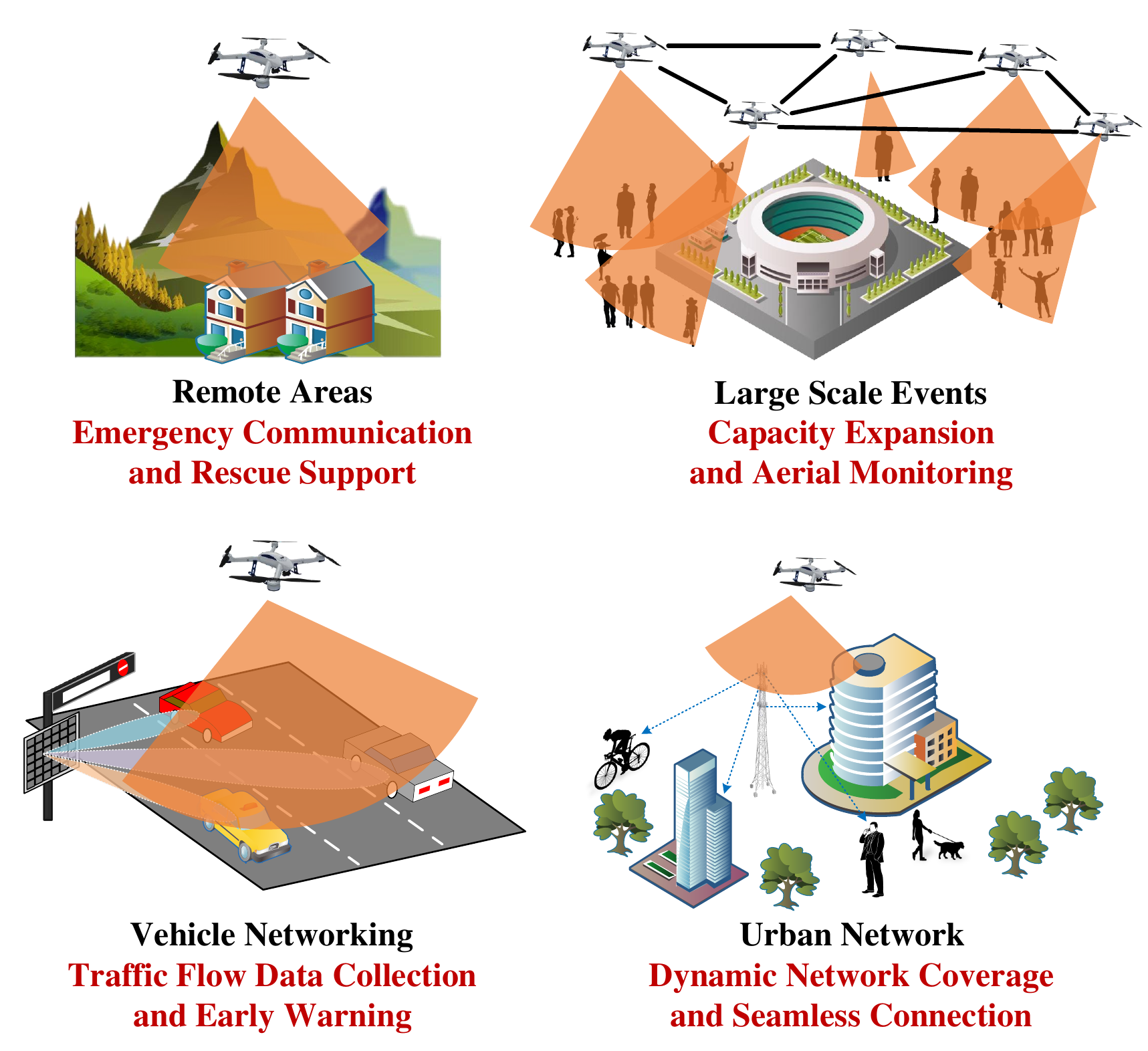}
		\caption{Typical UAV application scenarios in the LAE.}
		\label{fig1}
		\vspace{-5mm}
	\end{figure}
	
	Specifically, a typical diagram of UAV management in restricted areas is shown in Figure 2.
	To address the protection requirements of highly sensitive areas, the UAV security management system must not only accurately identify UAVs but also achieve high-precision localization and real-time dynamic data updating. On this basis, it further needs to implement effective disposal of targeted UAVs through countermeasures.
	Based on this, we divide the entire process of UAV management into identification, localization, and countermeasures\footnote{Although some devices can simultaneously achieve identification and localization of UAVs, there exist fundamental differences between the two in terms of implementation logic. Identification aims to determine the existence of UAVs, while localization focuses on accurately measuring their spatial coordinates. To avoid confusion, this paper maintains a clear distinction between these two key components.} \cite{liu2025toward, gao2024emerging}.
	The primary step is UAV identification.
	Based on the core sensing principles, UAV identification can be categorized into several approaches.
	UAV identification based on radio frequency (RF) signals is widely adopted.
	However, practical scenarios show that these signals exhibit time-frequency domain dynamic characteristics, with their modulation methods also evolving toward higher complexity \cite{10829621}.
	Another widely adopted solution is radar, which emits electromagnetic waves and receives target echoes for UAV detection \cite{8008140,9519838}.
	However, its limitations lie in the fact that the echo signals of small UAVs are excessively weak and prone to being submerged by background noise \cite{11130553, 7508914}.
	Opto-electronic (OE) sensors, meanwhile, are well-suited for detecting small, fast-moving targets at low altitudes \cite{9694151,4161589}. Nevertheless, their performance is highly dependent on environmental conditions.
	Acoustic UAV identification features low deployment costs and easy scalability \cite{10933667}. However, it is also susceptible to environmental noise interference and has a limited detection range \cite{8950116}.
	To address the inherent limitations of UAV unimodal identification, proven technological evolution experience from the autonomous driving \cite{10316635} can be drawn on to promote the transformation of the identification system towards a multi-sensor fusion architecture \cite{10138659}.
	Building on this, artificial intelligence (AI) technology can be introduced to fuse data from multiple sensors and extract spatiotemporal, spectral, and image features of UAVs, and thereby enhance the UAV identification performance.

	\begin{figure}[htbp]
		\centering
		\includegraphics[width=1\linewidth]{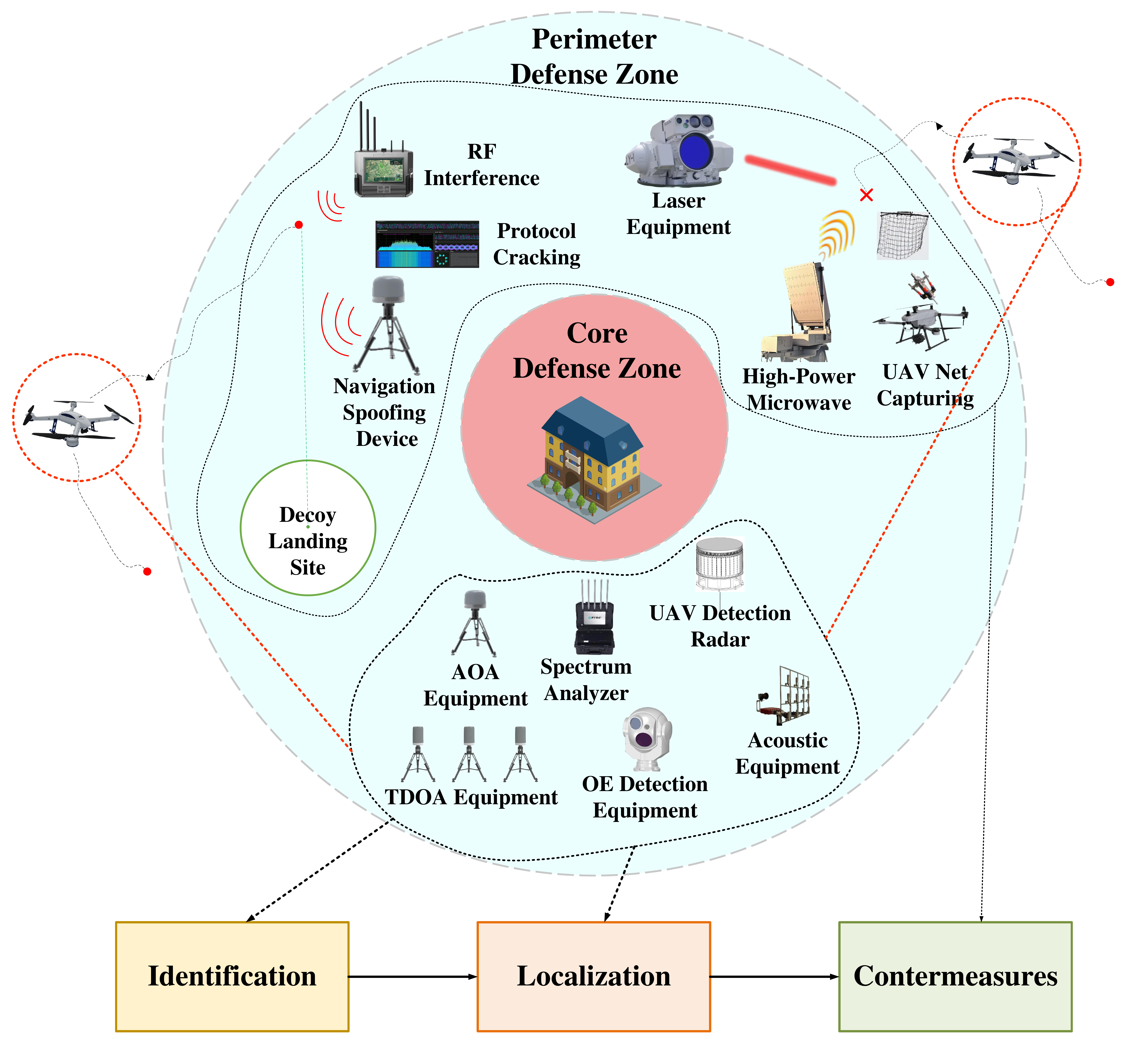}
		\caption{Diagram of UAV management in restricted areas.}
		\vspace{-4mm}
	\end{figure}
	
	\begin{figure*}[htbp]
		\centering
		\includegraphics[width=1\linewidth]{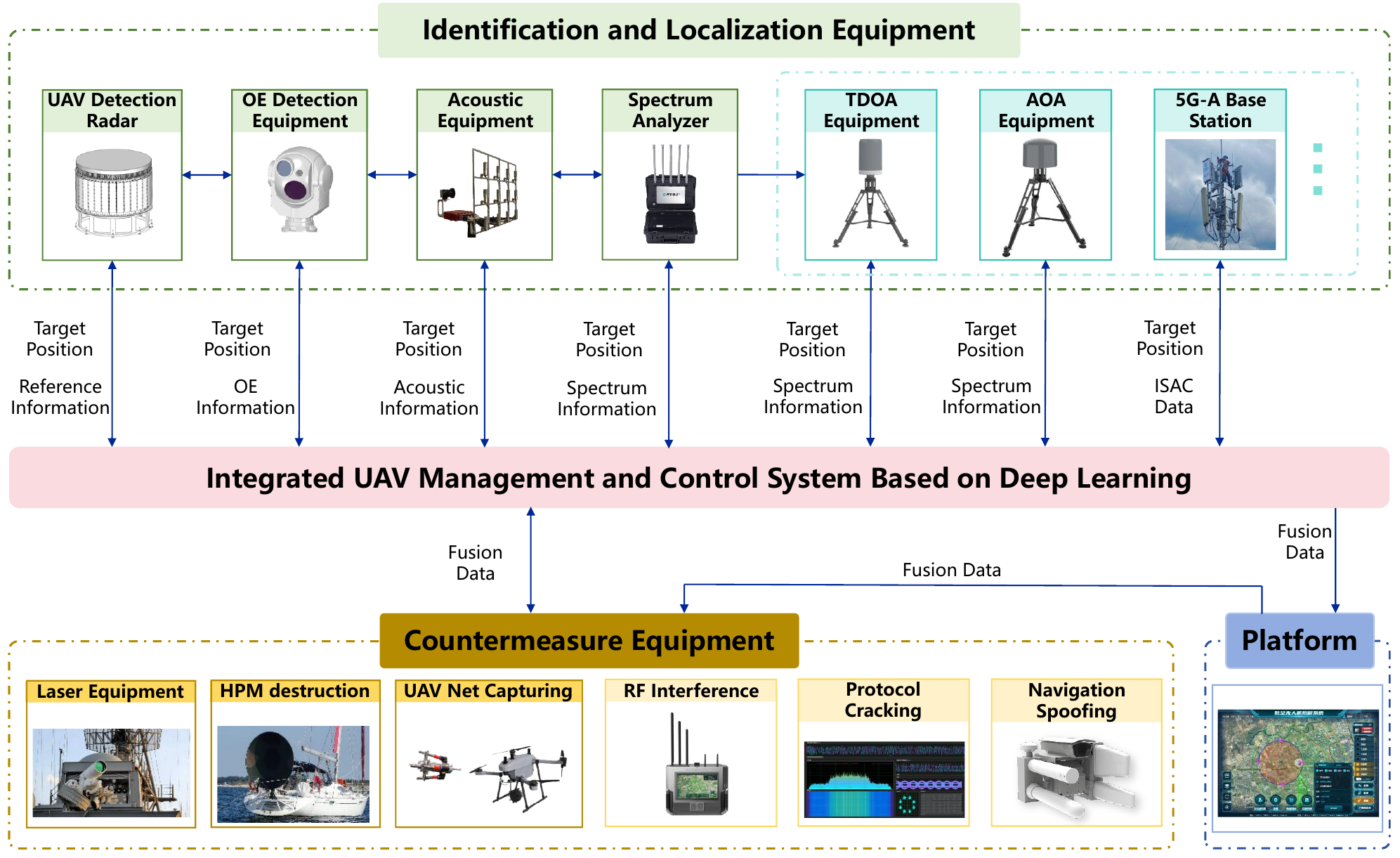}
		\caption{Integrated UAV management system based on deep learning.}
		\vspace{-4mm}
	\end{figure*}
	
	After identification, high-precision localization becomes the core link in UAV management \cite{8337900}.
	Currently, UAV localization technologies include time difference of arrival (TDOA) localization \cite{9748989}, angle of arrival (AOA) localization \cite{9146923}, ground radar localization \cite{10534287}, and so on.
	Additionally, by leveraging ground-based fifth-generation advanced (5G-A) integrated sensing and communications (ISAC) base station resources \cite{10064000}, the system can not only fulfill routine communication tasks but also achieve high-precision UAV identification and tracking.
	Similarly, each of these technologies demonstrates unique advantages in practical applications while also having certain limitations.
	Against this backdrop, AI technology can establish an intelligent localization system by fusing multi-sensor data and optimizing dynamic decision-making \cite{10172187}.
	Through the integration of a space-air-ground system, combined with AI algorithms, improvements in localization accuracy and coverage range can be achieved.
	
	Building upon these identification and localization capabilities, effective countermeasures are the critical step for UAV management.
	UAV countermeasures include two types of approaches, namely physical destruction (hard kill) and interference blocking (soft kill).
	Hard kill approaches include laser weapon destruction, high-power microwave (HPM) destruction \cite{9839577}, UAV net capture \cite{10323488}, and so on.
	Soft kill approaches include UAV RF jamming, navigation spoofing, protocol cracking, and so on \cite{7988845}. Although these two types of countermeasures can fulfill the basic objectives of UAV management, they still exhibit constrained response efficiency. Luckily, the integration of AI has yielded a critical breakthrough, enabling significantly enhanced countermeasure efficiency.
	Furthermore, AI can realize the allocation of countermeasure resources, allowing the system to determine the priority of countermeasure approaches.
	
	In this paper, we analyze and summarize the AI-based UAV management system capable of efficiently coordinating an integrated framework that combines identification, localization, and countermeasure functions, as shown in Figure 3.
	Throughout the entire UAV management process, AI breaks through the limitations by means of multi-sensor data fusion, dynamic decision optimization, and autonomous allocation of countermeasure resources, which significantly improve UAV identification accuracy, localization performance, and countermeasure effectiveness.
	
	\section{UAV Intelligent Identification: Fusion of Multi-Sensor Data and Deep Learning}
	With the growth in the number of UAVs and the gradual diversification of their application scenarios, the pressure on UAV safety supervision has been increasing. Therefore, UAV identification, as the core support for solving the dilemma of low-altitude supervision, is realizing an upgrade by means of multi-sensor data fusion and AI algorithm iteration.
	In this section, we focus on exploring the integration of UAV RF signal identification and AI technologies, while analyzing the comprehensive identification approaches based on multimodal multi-sensor data.
	
	\subsection{Deep Learning-based RF Signal Identification}
	UAV RF signal types include fixed-frequency periodic pulse signals, fixed-frequency continuous signals, narrowband frequency-hopping signals, and so on. Modulation methods exhibit highly composite characteristics, covering modulation modes formed by the combination of multiple modulation approaches \cite{s22083072}.
	Figure 4 presents the signal diagrams of two typical UAVs. As can be seen, signals of the two different UAVs exhibit significant differences in the time domain, frequency domain, and time-frequency domain.
	And such differences are precisely the key to distinguishing between UAVs.
	Nevertheless, UAV RF signals still exhibit several common characteristics, which are specifically manifested as follows.
	
	\begin{itemize}
		\item Most UAV signals are narrowband frequency-hopping signals, whose frequency point distribution is related to the type of data link \cite{10.1002}. Moreover, time-domain signals of different data link types also differ in terms of duration and signal duty cycle. Therefore, the characteristics can be used to distinguish between different types of UAV frequency-hopping signals.
		
		\item Specific types of UAV signals exhibit pulse emission characteristics, with highly stable emission cycles and duty cycles. Such periodic characteristics can serve as key discriminative criteria for signal detection \cite{9010185}.
		
	\end{itemize}
	\begin{figure}[htbp]
		\centering
		\includegraphics[width=1\linewidth]{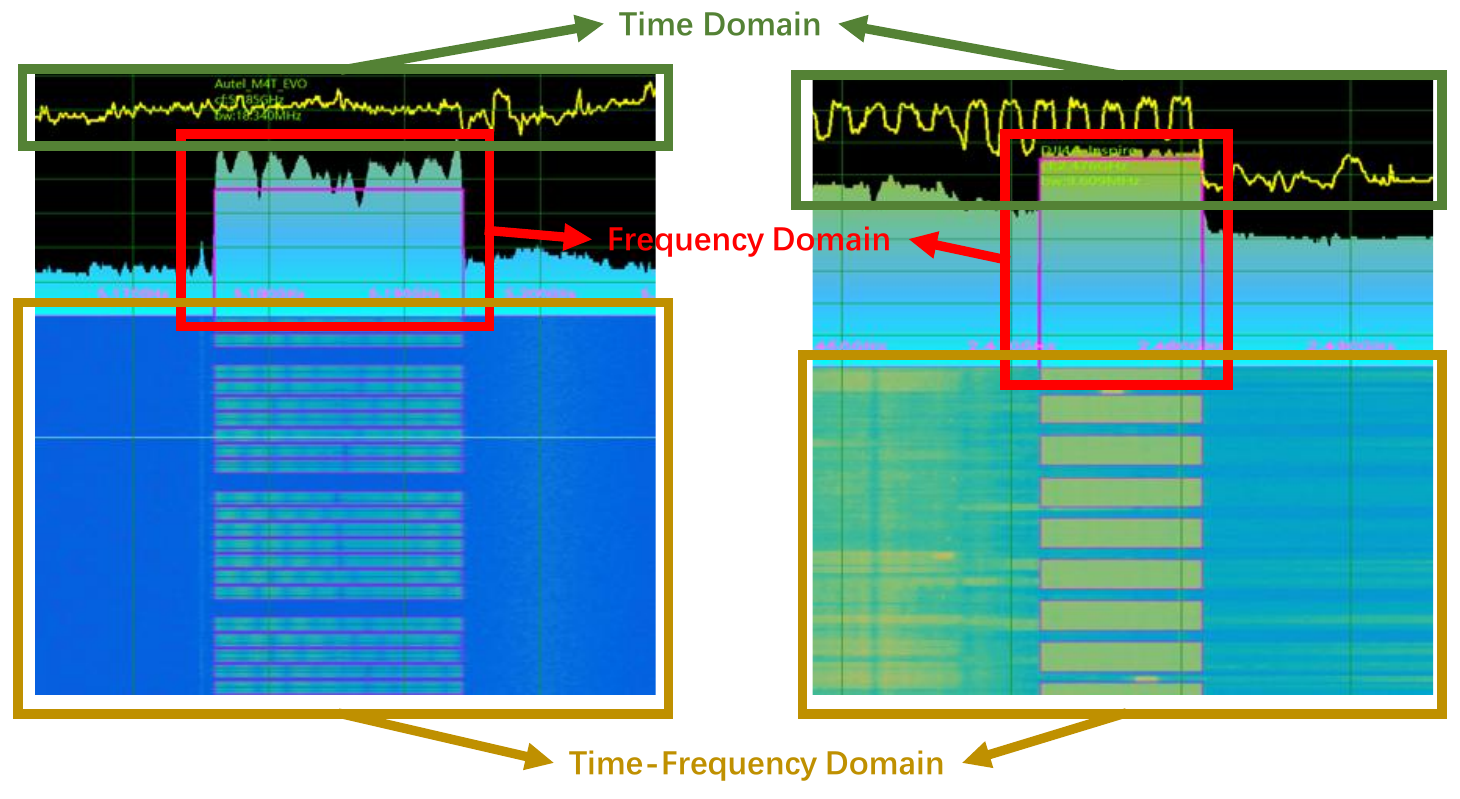}
		\caption{The comparison of RF signals between DJI Inspire 2 and Autel Robotics EVO Max 4T in the time domain, frequency domain, and time-frequency domain. }
		\vspace{-4mm}
	\end{figure}
	
	\begin{figure}[htbp]
		\centering
		\includegraphics[width=1\linewidth]{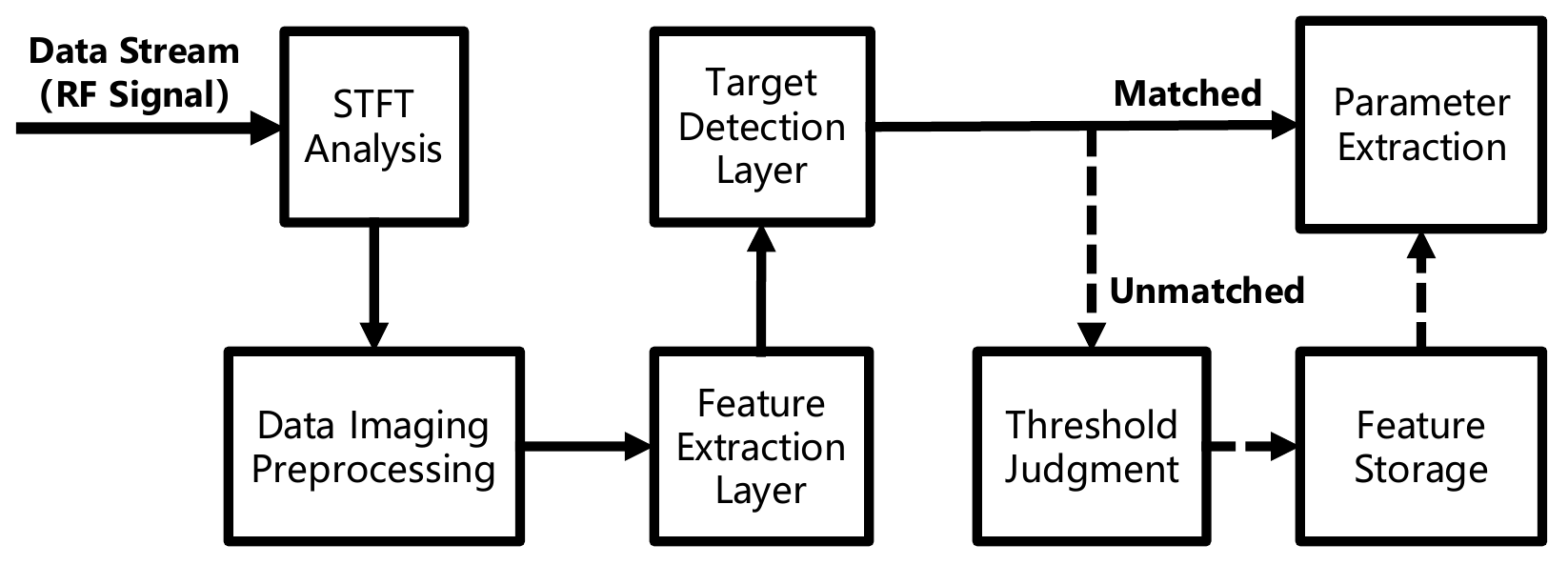}
		\caption{The flowchart of AI-based UAV RF signal identification.}
		\vspace{-4mm}
	\end{figure}
	
	However, the complex electromagnetic environment poses challenges to UAV RF signal identification. Across multiple dimensions such as the time domain, frequency domain, energy domain, and spatial domain, there exists a dense distribution of electromagnetic signals with diverse patterns and dynamic randomness, which severely undermines UAV RF identification performance.
	To address the aforementioned challenges, an AI-based signal identification solution is proposed, with its core workflow as illustrated in Figure 5.
	After being processed by short-time Fourier transform (STFT), the data stream is converted into a time-frequency map, where the vertical axis represents time, the horizontal axis represents frequency, and the color of pixels corresponds to the energy intensity \cite{9509437}.
	Then, some image preprocessing operations such as normalization are performed, followed by input into the deep learning model, which learns key features ranging from simple edges and textures to complex pattern structures, and encodes these features into high-dimensional feature vectors \cite{1004223,8846214}.
	Additionally, a pre-constructed UAV database is involved, whose core function is to store features of various known UAVs. These features can be used to classify the UAV corresponding to the current data stream \cite{9600566}.
	If the current data stream can be successfully matched with an existing UAV model in the database, signal processing can be performed based on the known attributes of this UAV model.
	Thus, core UAV parameters can be accurately extracted, providing data support for subsequent localization and countermeasures.
	In cases where the data stream fails to match any existing UAV model in the database, it is necessary to first conduct validity verification via preset thresholds. After confirming that the data stream contains a valid UAV RF signal, the unique features and key parameters of this unknown signal are further extracted, marked as a new UAV model, and incorporated into the database \cite{11085123}.
	This scheme has undergone preliminary verification.
	For instance, \cite{10942274} proposes an ASPConv-ResNet34 model for UAV identification, achieving an average test accuracy of 99.97\%, outperforming non-deep learning approaches by a significant margin.

	\begin{figure}[htbp]
		\centering
		\includegraphics[width=1\linewidth]{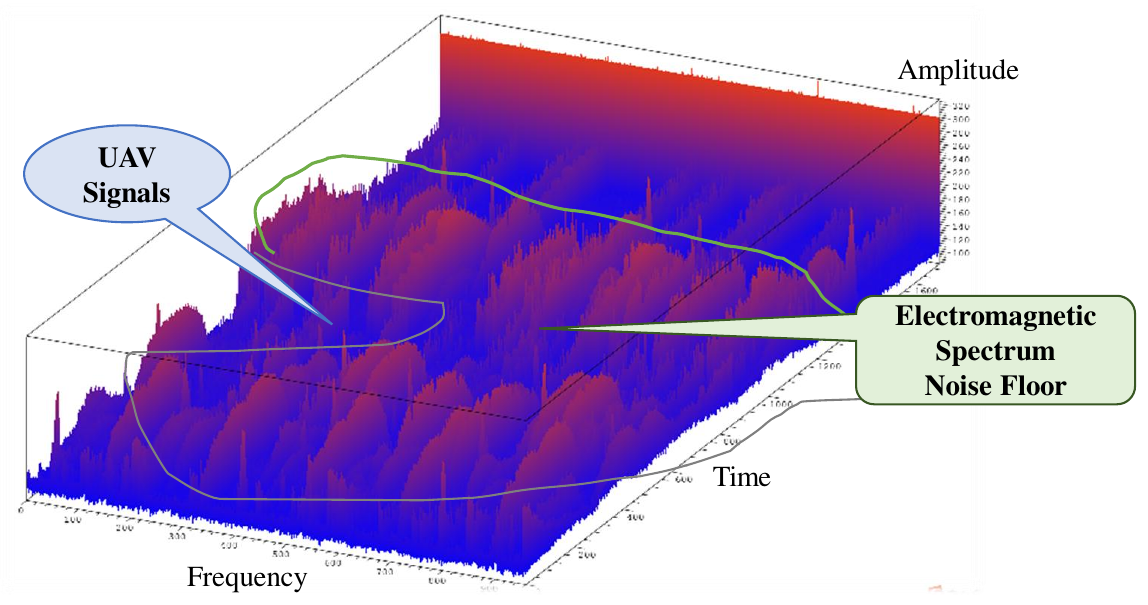}
		\caption{UAV signals and electromagnetic spectrum noise floor in low-altitude environments.}
		\vspace{-4mm}
	\end{figure}
	
	\begin{table*}
		\centering
		\caption{Comparison of UAV Identification \& Localization Approaches.}
		\footnotesize
		\renewcommand{\arraystretch}{1.3}
		\renewcommand{\tabularxcolumn}[1]{m{#1}}
		\newcolumntype{Y}{>{\centering\arraybackslash}X}
		\begin{tabularx}{\textwidth}{@{} c Y Y Y Y @{}}
			\toprule
			\textbf{Approach} & \textbf{Range} & \textbf{Precision} & \textbf{Limitation} & \textbf{Application} \\
			\midrule
			\textbf{RF (TDOA/AOA)}
			& $\geq$ 3 km
			& $\leq$ 30 m
			& Silent UAVs undetectable
			& Urban grid monitoring; Multi-site localization \\
			\midrule
			\textbf{RF (Remote ID)}
			& $\leq$ 2 km (Broadcast)
			& GNSS dependent
			& Relies on active cooperation
			& Cooperative target authentication; Whitelist management\\
			\midrule
			\textbf{Radar}
			& 5--10 km
			& $\leq$ 10  m
			& Weak on small/slow targets; High false alarm rate
			& Large airports; Open areas \\
			\midrule
			\textbf{OE}
			& 1--2 km
			& High (Visual)
			& Weather/Light sensitive
			& Forensics; Visual confirmation \\
			\midrule
			\textbf{5G-A (ISAC)}
			& 1.2--1.5 km (Single)
			& $\leq$ 10 m
			& Early commercial stage
			& Core protected zones; Continuous networking \\
			\bottomrule
		\end{tabularx}
	\end{table*}
	
	To further mitigate signal interference in complex electromagnetic environments, high-precision electromagnetic environment modeling can be developed.
	As shown in Figure 6, the interference from other electromagnetic signals in the scenario can be extremely severe \cite{8888398}.
	By integrating the electromagnetic spectrum noise floor, signal time-frequency characteristics, statistical distribution properties, and domain prior knowledge, the complex electromagnetic behaviors of various radiation sources can be accurately simulated \cite{8464269}.
	Additionally, incorporating the constructed electromagnetic environment as noise or disturbance input into the training process can effectively enhance the robustness and generalization capability of the models under real electromagnetic interference.

	\begin{figure}[htbp]
		\centering
		\begin{subfigure}{0.48\linewidth}
			\centering
			\includegraphics[width=\linewidth, keepaspectratio]{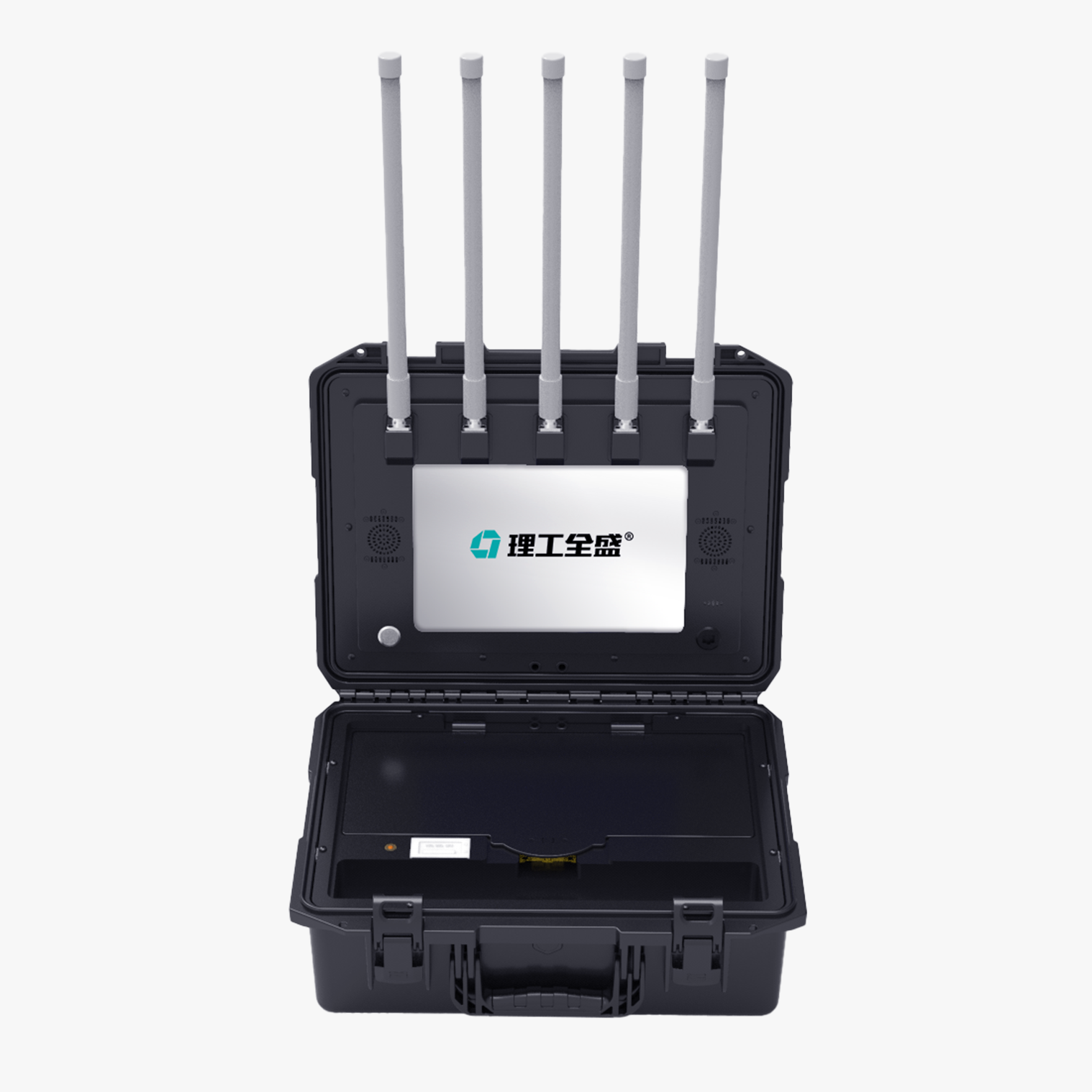}
			\caption{} 
			\label{fig:dm}
		\end{subfigure}
		\hfill
		\begin{subfigure}{0.48\linewidth}
			\centering
			\includegraphics[width=\linewidth, keepaspectratio]{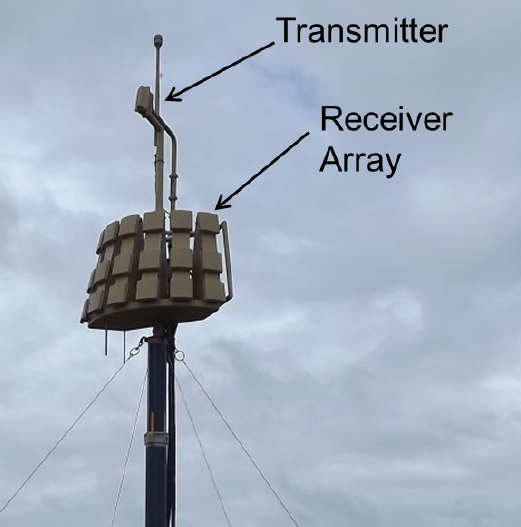}
			\caption{} 
			\label{fig:radar}
		\end{subfigure}
		
		\vspace{1ex} 
		
		\begin{subfigure}{0.48\linewidth}
			\centering
			\includegraphics[width=\linewidth, keepaspectratio]{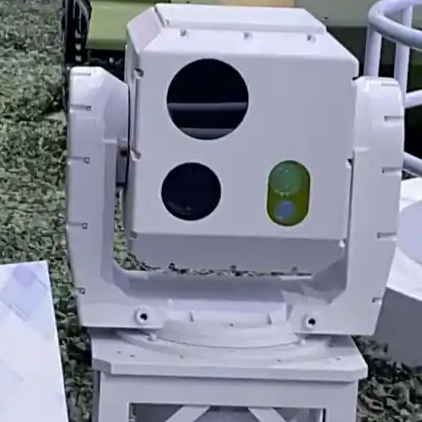}
			\caption{} 
			\label{fig:opto}
		\end{subfigure}
		\hfill
		\begin{subfigure}{0.48\linewidth}
			\centering
			\includegraphics[width=\linewidth, keepaspectratio]{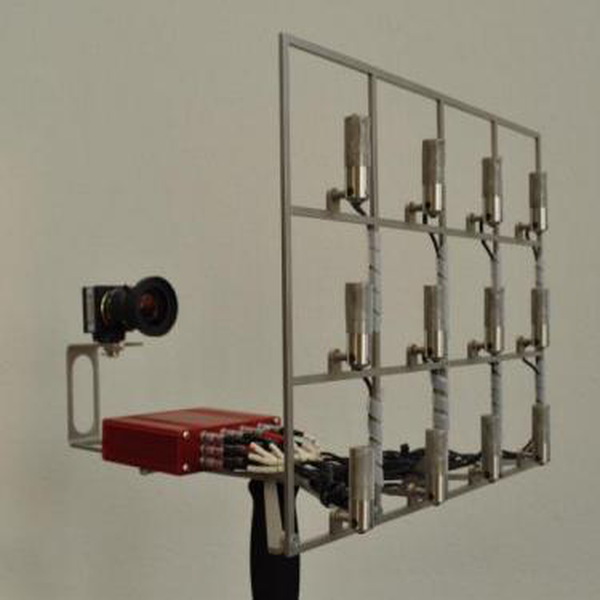}
			\caption{} 
			\label{fig:mic}
		\end{subfigure}
		
		\caption{Several different UAV identification technologies: (a) Spectrum analyzer; (b) Radar \cite{7346268}; (c) OE devices; (d) Acoustic sensors.}
		\vspace{-4mm}
	\end{figure}
	
	\subsection{Multi-Sensor Data Fusion Identification}
	Beyond RF signal identification, there are a variety of identification approaches for UAVs, as shown in Figure 7.
	Specifically, radar possesses long-range detection capabilities and is not affected by lighting or weather conditions.
	In terms of performance, the radar attains an identification probability of at least 85\%, coupled with a false alarm rate of no more than 5\% \cite{9409707}.
	However, it exhibits low sensitivity to low-altitude, slow-moving, small UAVs and struggles to distinguish between hovering UAVs and static objects \cite{10375084}.
	OE sensors can provide high-resolution images for extracting the shape, texture, and thermal radiation features of UAVs \cite{9694151}. However, they are highly affected by weather conditions and involve high computational complexity.
	Acoustic sensors can distinguish UAVs from non-target objects such as birds by capturing the noise harmonics and transient characteristics of UAV propellers \cite{s20143923}, which feature a fully passive identification capability, and are sensitive to low-altitude, slow-moving targets. Nevertheless, they suffer from limited detection range, are susceptible to interference from environmental noise, and struggle to differentiate individual UAVs within swarms.
	
	Given the inherent performance limitations of individual UAV identification technologies in complex environments, multi-source data fusion has emerged as a key approach to enhancing system robustness and accuracy.
	By integrating multimodal multi-sensor data, extracting features from different modalities, conducting collaborative identification based on various sensors, and implementing adaptive optimization according to scenarios, the robustness of UAV identification systems can be enhanced \cite{9774975}.
	Specifically, data fusion includes data-level, feature-level, and decision-level fusion.
	Data-level fusion directly fuses data collected by multiple heterogeneous sensors with minimal or no preprocessing to form a comprehensive composite dataset \cite{10.1117/12.2306148}.
	A common strategy is to convert heterogeneous data into a unified representation. Subsequently, feature extraction and target identification are performed. This fusion can theoretically achieve optimal identification accuracy. However, it imposes high requirements on communication bandwidth, computational resources, and so on.
	Feature-level fusion independently processes data and extracts features for each heterogeneous sensor, e.g., modulation features, radar parameter features, OE texture features, and so on \cite{10453239}.
	These vectors are then concatenated into a unified high-dimensional joint feature for identification. Note that feature extraction may lead to information loss, and feature selection is crucial to the final performance.
	Decision-level fusion treats each sensor as an independent system, completing the full process from preprocessing and feature extraction to preliminary local decision-making \cite{11218172}. Specific fusion rules are subsequently adopted to integrate local decisions into a globally optimal joint decision, which offers high flexibility and fault tolerance. However, prior local decision-making may result in severe information loss.
	By effectively combining these fusions, the UAV management system can leverage the respective strengths of diverse systems, mitigating the inherent limitations of any single modality.
	
	\section{High-precision Localization and Real-time Tracking of UAVs}
	After achieving UAV identification, the system needs to further perform high-precision localization and continuous tracking of UAVs.
	Different identification and localization approaches exhibit distinct characteristics of their own, as shown in Table 1.
	Among these, some approaches do not require devices to actively transmit signals, thus making them suitable for scenarios with high requirements for concealment.
	TDOA localization calculates the position of the UAV by measuring the time difference of signal arrival at multiple devices, which reduces system complexity \cite{9748989}.
	Notably, it requires a network of at least three time-synchronized stations and delivers high positioning accuracy, making it well-suited for large-scale low-altitude defense scenarios.
	AOA localization determines the position of UAVs using the azimuth and elevation angles of the signal arriving at devices \cite{9146923}.
	Similarly, to achieve effective localization, AOA localization requires a minimum of two stations \cite{7782329}.
	Remote Identification (Remote ID) is a mandatory technical standard that requires UAVs to proactively broadcast critical information such as unique ID, position, altitude, and speed during flight. Based on the transmitted data, UAVs can be identified and localized in real time, thereby enhancing airspace safety \cite{9861637}.
	Furthermore, some active UAV localization approaches, e.g., ground-based radar, can also achieve UAV localization after identification.
	
	\begin{figure}[htbp]
		\centering
		\begin{subfigure}{0.48\linewidth}
			\centering
			\includegraphics[width=\linewidth, keepaspectratio]{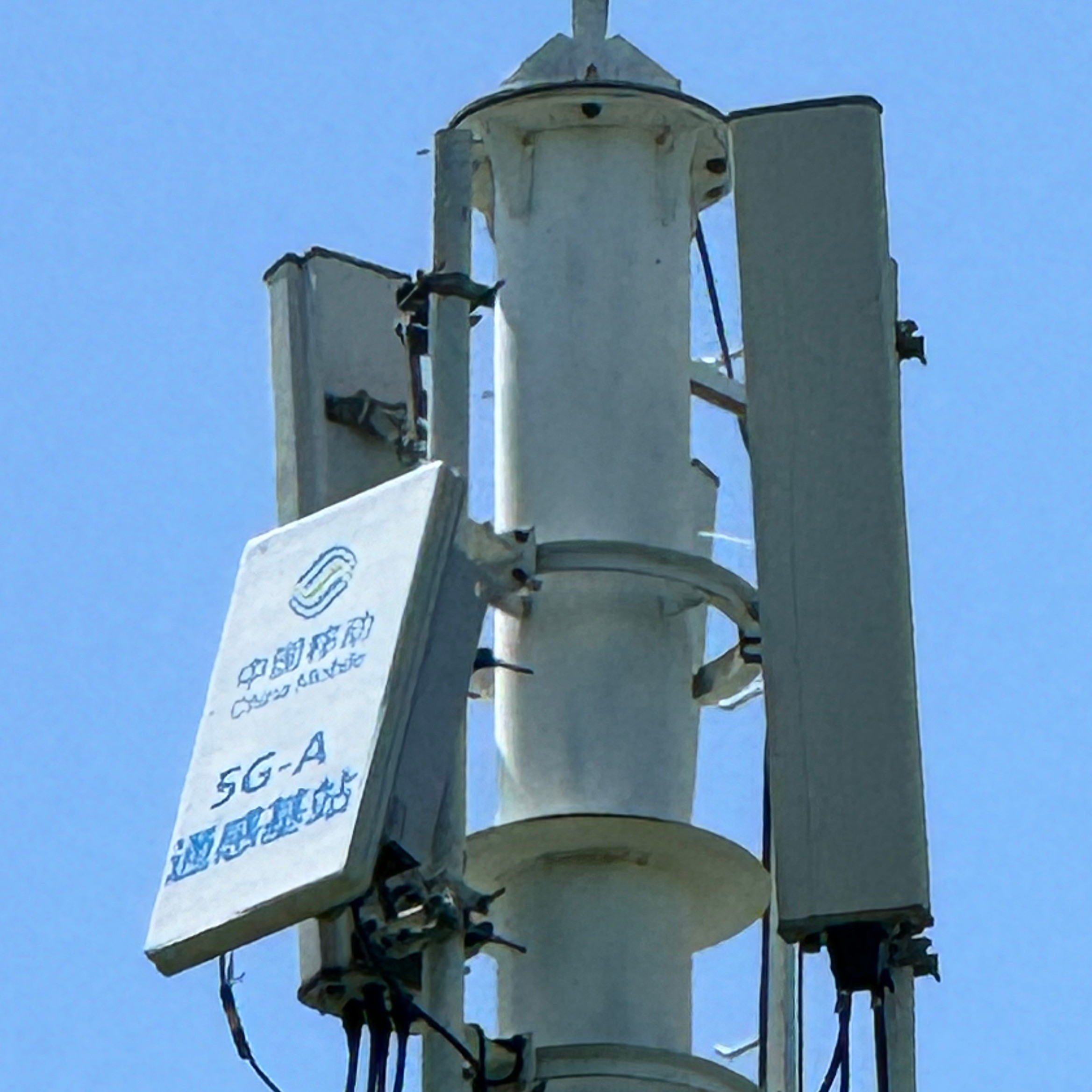}
			\caption{}
		\end{subfigure}
		\hfill
		\begin{subfigure}{0.48\linewidth}
			\centering
			\includegraphics[width=\linewidth, keepaspectratio]{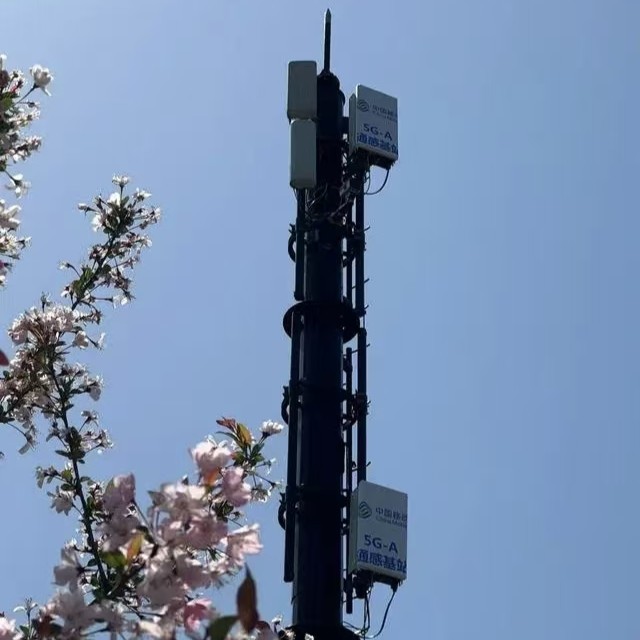}
			\caption{}
		\end{subfigure}
		
		\vspace{0.5em}
		
		\begin{subfigure}{0.48\linewidth}
			\centering
			\includegraphics[width=\linewidth, keepaspectratio]{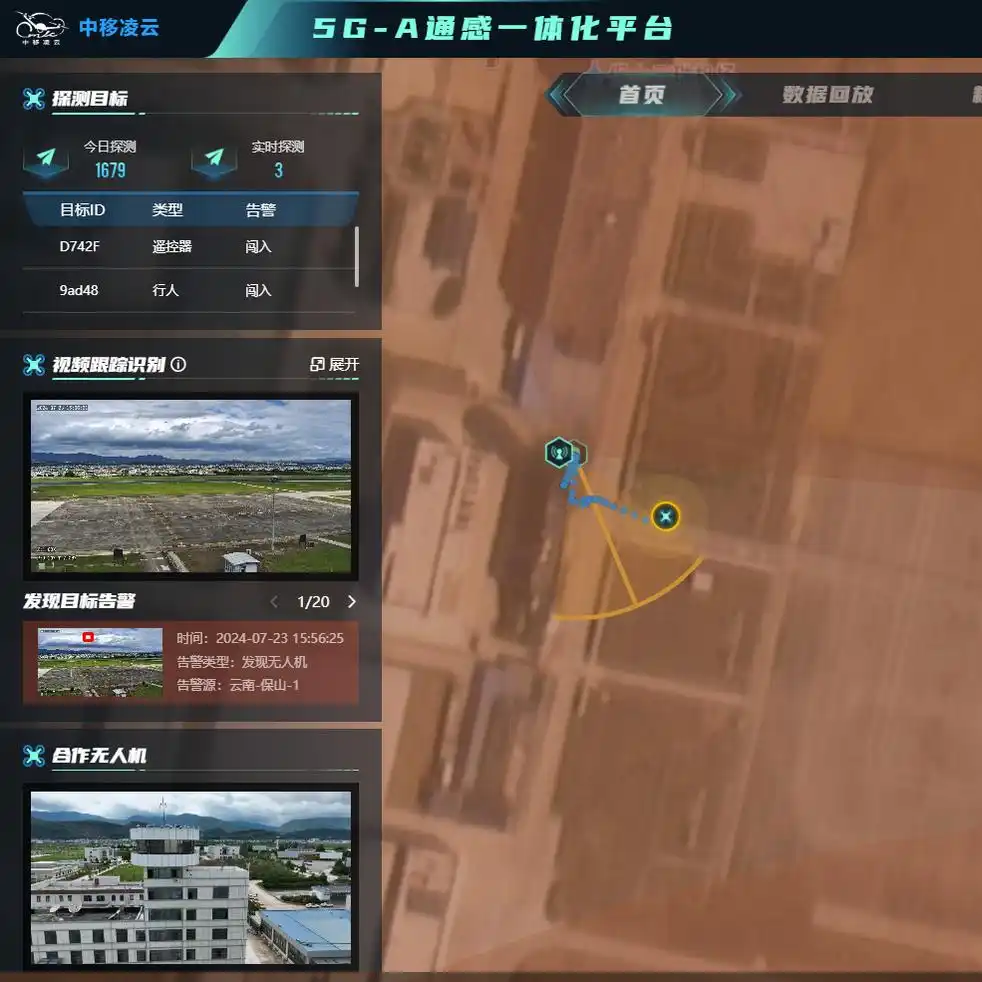}
			\caption{}
		\end{subfigure}
		\hfill
		\begin{subfigure}{0.48\linewidth}
			\centering
			\includegraphics[width=\linewidth, keepaspectratio]{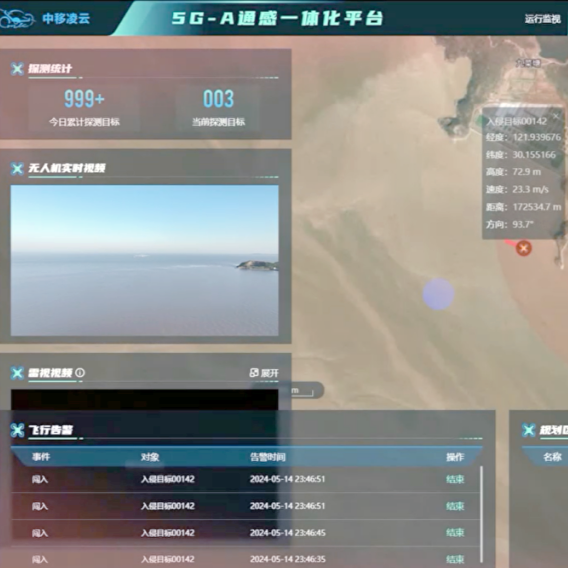}
			\caption{}
		\end{subfigure}
		
		\caption{The practical applications of 5G-A ISAC: (a)\&(b) Equipment deployment; (c) 5G-A low-altitude ISAC at Baoshan Airport in Yunnan; (d) 5G-A low-altitude ISAC at Zhoushan cross-sea route in Zhejiang.}
		\vspace{-4mm}
	\end{figure}
	
	In recent studies, ISAC has demonstrated substantial application potential in low-altitude UAV management by sharing hardware and spectrum resources \cite{9456851,9898900,10886954}.
	As illustrated in Figure 8, 5G-A ISAC equipment has been deployed and validated in real-world scenarios \cite{9849060,10195164,7001610}, which fully leverages existing base station sites, transmission networks, and operation and maintenance systems.
	A typical 5G-A ISAC system can operate in the 4.9 GHz and 26 GHz frequency bands, achieving an identification probability of $\geq$ 95\% while maintaining a false alarm rate of $\leq$ 5\%. It is capable of providing a localization accuracy of $\leq$ 10 meters.
	However, it still confronts several challenges in advancing toward large-scale application \cite{9174777,9527086}. On the one hand, key performance indicators such as sensing accuracy require sufficient validation across a broader spectrum of real-world scenarios. On the other hand, compared with dedicated low-altitude radar systems, the sensing coverage of current single ISAC base stations is still limited, necessitating multi-station cooperative networking to enhance overall performance.
	
	Similar to UAV identification, multi-sensor data fusion and cooperative positioning are key to enhancing the accuracy and robustness of UAV localization and tracking systems \cite{10653358}.
	By integrating multi-sensor data and combining multimodal algorithms, a space-air-ground cooperative localization system with full-domain coverage and real-time response capabilities can be built \cite{guo2023ergodic}, which relies on the integration of space-based navigation and ground-based systems to provide support for UAV localization and countermeasures.
	Space-based augmentation is typically composed of near-space aerostats (such as stratospheric airships and high-altitude balloons) and UAV platforms \cite{10770537}.
	Through collaborative networking, these platforms form an integrated network covering from the ground to low altitudes, which is used to capture and identify UAV signals, achieving complementary coverage in terms of range.
	In terms of ground deployment, by systematically integrating the relevant technologies and upgrading infrastructure (e.g., 5G-A ISAC base stations), efficient and large-scale coverage of communication networks can be achieved at a relatively low cost.
	
	\begin{figure}[htbp]
		\centering
		\includegraphics[width=1\linewidth]{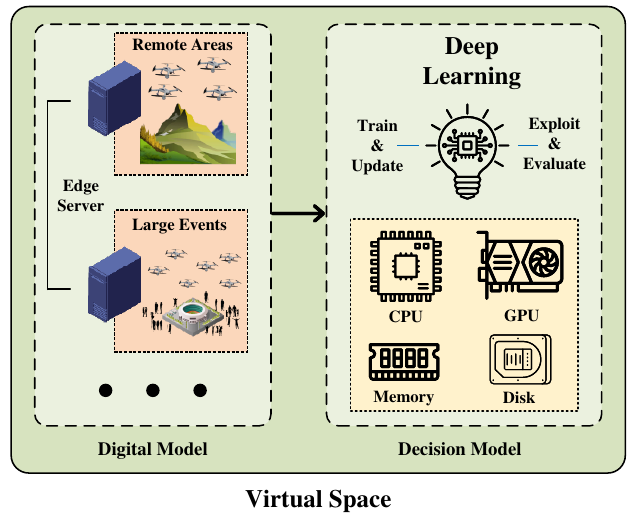}
		\caption{Overview of the digital twin-based intelligent cooperation framework of UAV swarm.}
		\vspace{-2mm}
	\end{figure}
	
	\begin{figure*}[htbp]
		\centering
		\begin{subfigure}{0.315\linewidth}
			\centering
			\includegraphics[width=\linewidth, keepaspectratio]{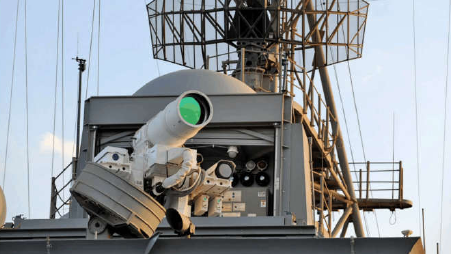}
			\caption{}
		\end{subfigure}
		\hfill
		\begin{subfigure}{0.278\linewidth}
			\centering
			\includegraphics[width=\linewidth, keepaspectratio]{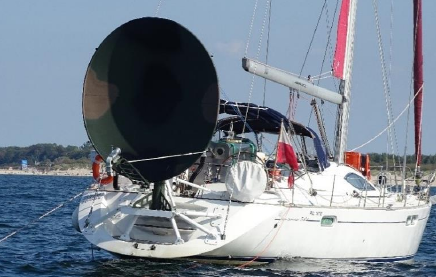}
			\caption{}
		\end{subfigure}
		\hfill
		\begin{subfigure}{0.30\linewidth}
			\centering
			\includegraphics[width=\linewidth, keepaspectratio]{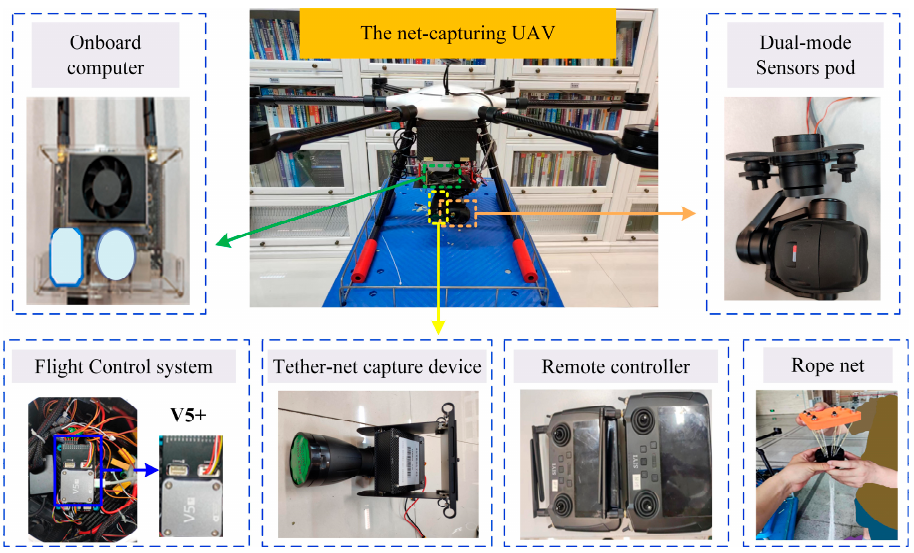}
			\caption{}
		\end{subfigure}
		
		\vspace{-4mm}
		\caption{Typical UAV hard-kill approaches: (a) Laser weapons \cite{10318249}; (b) HPM \cite{10644353}; (c) A net-capturing UAV \cite{drones8100573}.}
		\vspace{-2mm}
	\end{figure*}
	
	\begin{figure*}[htbp]
		\centering
		\begin{subfigure}{0.3\linewidth}
			\centering
			\includegraphics[width=\linewidth, keepaspectratio]{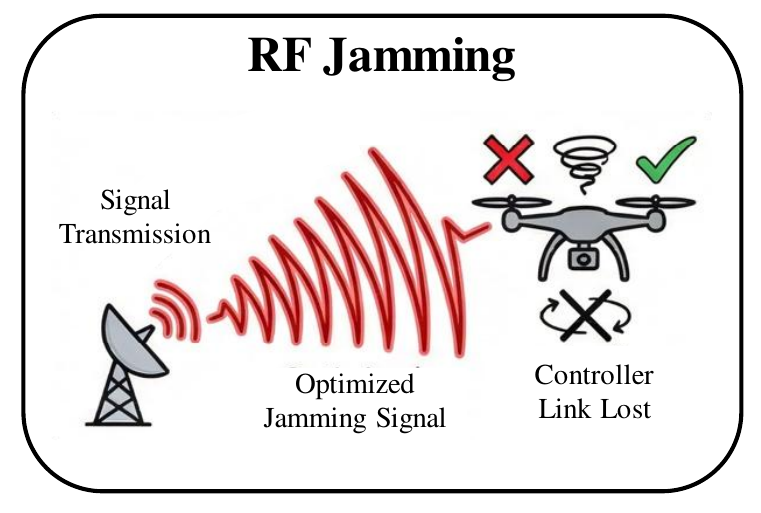}
			\caption{}
		\end{subfigure}
		\hfill
		\begin{subfigure}{0.3\linewidth}
			\centering
			\includegraphics[width=\linewidth, keepaspectratio]{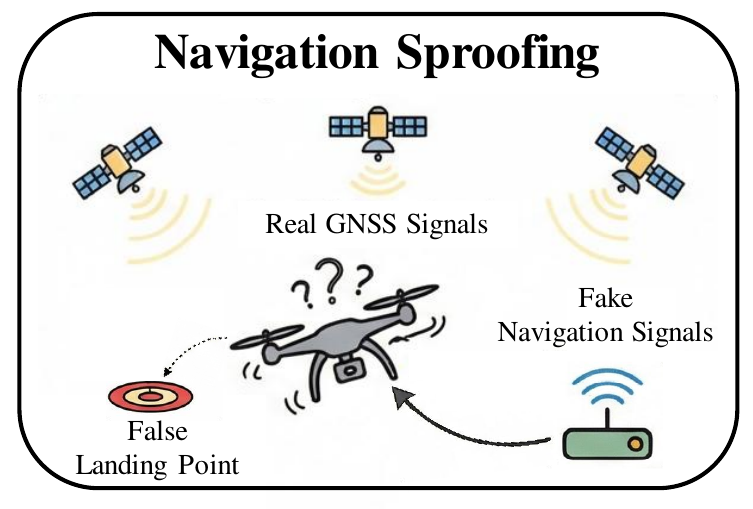}
			\caption{}
		\end{subfigure}
		\hfill
		\begin{subfigure}{0.3\linewidth}
			\centering
			\includegraphics[width=\linewidth, keepaspectratio]{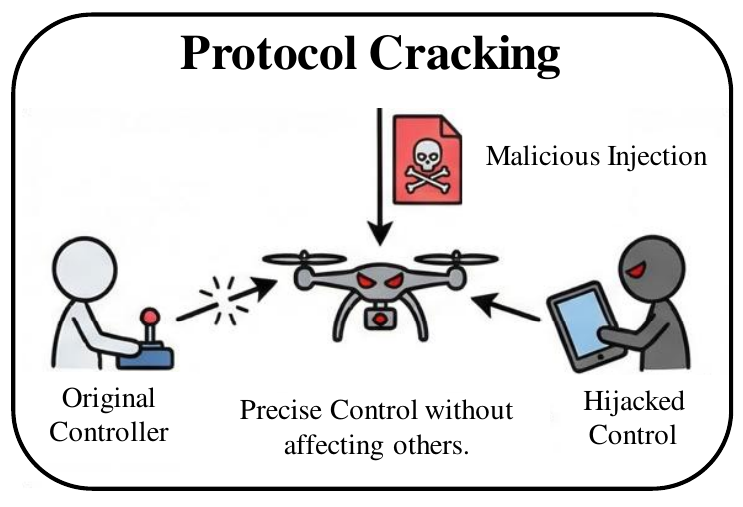}
			\caption{}
		\end{subfigure}
		
		\vspace{-4mm}
		\caption{Typical UAV soft-kill approaches: (a) RF jamming; (b) Navigation spoofing; (c) Protocol cracking.}
		\vspace{-4mm}
	\end{figure*}
	
	In the space-air-ground integrated cooperative localization system, AI serves as the core for processing multi-sensor data and for optimizing UAV tracking and trajectory prediction \cite{9504602,9615243}.
	Specifically, the deep learning model for UAV localization can further optimize the tracking process by leveraging its pre-trained spatial feature extraction capabilities to quickly associate the spatiotemporal consistency of multi-source data, thereby reducing the errors incurred by traditional algorithms during data alignment. Meanwhile, it adaptively adjusts the fusion strategy based on dynamic environmental conditions, providing higher-quality preprocessed data for subsequent trajectory generation and enabling accurate, stable, and reliable tracking and prediction of UAV flight trajectories \cite{10648926}.
	For instance, Transformer \cite{Yu_2023_CVPR} has demonstrated significant advantages in time-series data processing due to its strong long-sequence modeling capability and efficient parallel computing performance, and has gradually been applied to UAV tracking and trajectory prediction tasks.
	Additionally, empowered by AI, digital twin technology \cite{10329973} has achieved a qualitative leap, as shown in Figure 9.
	By integrating multi-sensor data and combining it with real-scene modeling, this technology can construct real-time environmental maps with semantic labels and generate a digital model of the airspace \cite{10190734,9263396}. Through correlation analysis within an AI fusion framework, it enables the continuous and accurate positioning of UAVs and also allows for the identification of key environmental elements using semantic labels.
	
	\begin{table*}
		\centering
		\caption{Performance comparison of UAV Countermeasures.}
		\label{tab:countermeasures}
		\footnotesize
		\renewcommand{\arraystretch}{1.5}
		
		\renewcommand{\tabularxcolumn}[1]{>{\centering\arraybackslash}m{#1}}
		
		\begin{tabularx}{\textwidth}{@{} c c c c c X X @{}}
			\toprule
			\textbf{Category} & \textbf{Approach} & \textbf{Attack Type} & \textbf{Attack Band} & \textbf{Range} & \textbf{Advantages} & \textbf{Limitations} \\
			\midrule
			
			\multirow{6}{*}[-0.5em]{\rotatebox{90}{\textbf{Hard Kill}}}
			& \makecell{\textbf{Laser}\\\textbf{Weapon}}
			& Optical
			& \makecell{Optical Band}
			& $\leq$ 2 km
			& High precision; Low marginal cost
			& Highly sensitive to weather; High system cost \\
			\cmidrule{2-7}
			
			& \textbf{HPM}
			& \makecell{Passive RF}
			& \makecell{Wideband Pulse}
			& 0.5--3 km
			& Area denial
			& Large physical footprint; Risk of collateral damage to electronics \\
			\cmidrule{2-7}
			
			& \textbf{Net Capture}
			& \makecell{Kinetic\\(Physical Net)}
			& N/A
			& $\leq$ 300 m
			& Minimal collateral risk
			& Single-target limitation; Difficult to aim \\
			\midrule
			
			\multirow{12}{*}[5em]{\rotatebox{90}{\textbf{Soft Kill}}}
			& \makecell{\textbf{RF Jamming}}
			& Passive RF
			& \makecell{433 MHz / 900 MHz\\1.2GHz / 1.5GHz (GNSS)\\2.4GHz / 5.8GHz}
			& \makecell{1--3 km (Omnidirectional)\\3--5 km (Directional)}
			& Effective against non-autonomous UAVs
			& High electromagnetic interference pollution \\
			\cmidrule{2-7}
			
			& \makecell{\textbf{Navigation}\\\textbf{Spoofing}}
			& \makecell{Passive RF}
			& \makecell{GPS L1 / L2\\BDS B1\\GLONASS}
			& $\geq$ 5 km
			& Can redirect or force-land UAVs
			& Ineffective against Inertial-based navigation\\
			\cmidrule{2-7}
			
			& \makecell{\textbf{Protocol}\\\textbf{Cracking}}
			& \makecell{Passive RF}
			& 2.4 GHz / 5.8 GHz
			& 3--5 km
			& Zero collateral interference
			& Ineffective against DIY UAVs \\
			\bottomrule
		\end{tabularx}
	\end{table*}
	
	\section{Deep Learning-Powered Hybrid Countermeasures for UAVs}
	UAV countermeasure approaches can be generally divided into physical destruction (hard-kill) and jamming-blocking (soft-kill), as shown in Table 2.
	The integration of AI can significantly enhance the performance of individual UAV countermeasure approaches.
	Furthermore, AI can assess the current threat level, available resource costs, and environmental constraints to autonomously determine the optimal interception strategy, thereby achieving the best possible outcome in various scenarios.
	
	\subsection{Deep Learning-Powered UAV Hard-Kill Approaches}
	Specifically, UAV hard-kill approaches are widely used in battlefield scenarios. Hard-kill technologies directly destroy or capture UAVs through physical means, mainly including laser weapons, HPM weapons, and net-capturing systems, as shown in Figure 10.
	Among these countermeasures, HPM weapons boast a countermeasure range of $\leq$ 3 km.
	Laser weapons rank second with a range of $\leq$ 2 km, while UAV net-capturing systems are typically limited to the operator's visual line of sight and conduct maneuverable interception through close-range proximity operations.
	
	On this basis, deep learning can improve the tracking capabilities of laser weapons, enabling early prediction of the trajectory of high-speed maneuvering targets and adjustment of beam direction \cite{karkadakattil2025ai}.
	The integration of deep learning with HPM weapons \cite{9839577} is mainly reflected in threat assessment and attack parameter optimization. Deep learning can real-time analyze the swarming patterns and behavioral characteristics of UAVs, and intelligently adjust the transmission power and action range of microwave weapons.
	The application of deep learning in net-capturing systems focuses on target positioning and interception path planning \cite{10323488}, which can accurately identify the model and attitude of the target UAV and predict its movement trajectory by leveraging the vision module mounted on the dual-mode sensors pod in conjunction with vision-based deep learning algorithms.
	Meanwhile, the onboard computer, relying on deep reinforcement learning algorithms, can generate efficient approaching paths and capture strategies for the net-capturing UAV, and then precisely execute these actions via the flight control system, thus significantly enhancing the success rate of intercepting the target UAV \cite{9530723}.
	Beyond the AI-enabled collaborative integration of multiple hard-kill approaches, the miniaturization and integration of weapon systems, as well as precision directed-energy emission technology, have also emerged as key research focuses \cite{wei2025uav}.
	
	\subsection{Deep Learning-Powered UAV Soft-Kill Approaches}
	However, in crowded scenarios with high human density, such as urban centers and venues for major events, hard-kill countermeasures may cause UAVs to lose control and crash, leading to serious secondary damage.
	Soft-kill technologies can disable UAVs without destroying the physical entity, which can significantly minimize collateral damage \cite{7988845}. Such technologies include RF jamming, navigation spoofing, and protocol cracking, as shown in Figure 11.
	These approaches typically have a countermeasure range exceeding 3 km and are primarily targeted at UAVs that actively transmit signals outward.
	
	The integration of deep learning technology has greatly enhanced the adaptive jamming and signal recognition capabilities of soft-kill systems.
	The combination of deep learning and RF jamming is mainly reflected in intelligent signal recognition and adaptive jamming strategies \cite{electronics11193025}.
	On this basis, generative adversarial networks can even generate optimized jamming signal patterns to implement the most effective jamming for specific UAVs while minimizing the impact on surrounding legitimate communications.
	Navigation spoofing transmits forged navigation signals to UAVs for luring them into designated locations or forcing them to land  \cite{10705273}. Due to the uncertainty in the movement characteristics of UAVs, existing approaches struggle to achieve precise control over the actual landing position of the UAV. Therefore, a deep learning-based progressive feedback navigation spoofing strategy can be adopted to implement a closed-loop behavioral response for navigation spoofing. By real-time updating spoofing tactics through environmental feedback, learning and predicting the navigation behavior patterns of UAVs, and generating highly deceptive fake navigation signals, this strategy can ultimately achieve accurate and controllable fixed-point forced landing.
	Protocol cracking is an emerging soft-kill technology that takes over the control of UAVs by cracking UAV communication protocols or implanting malicious code \cite{5207640}, which can achieve precise control of UAVs without affecting other devices. Sequence models based on long short-term memory networks can analyze the communication protocol patterns of UAVs and automatically identify protocol vulnerabilities and encryption weaknesses \cite{10000919}. Attack generation systems can generate effective control commands in real-time based on the learned protocol characteristics to hijack the control of UAVs.
	
	\begin{figure*}[htbp]
		\centering	\includegraphics[width=0.8\linewidth]{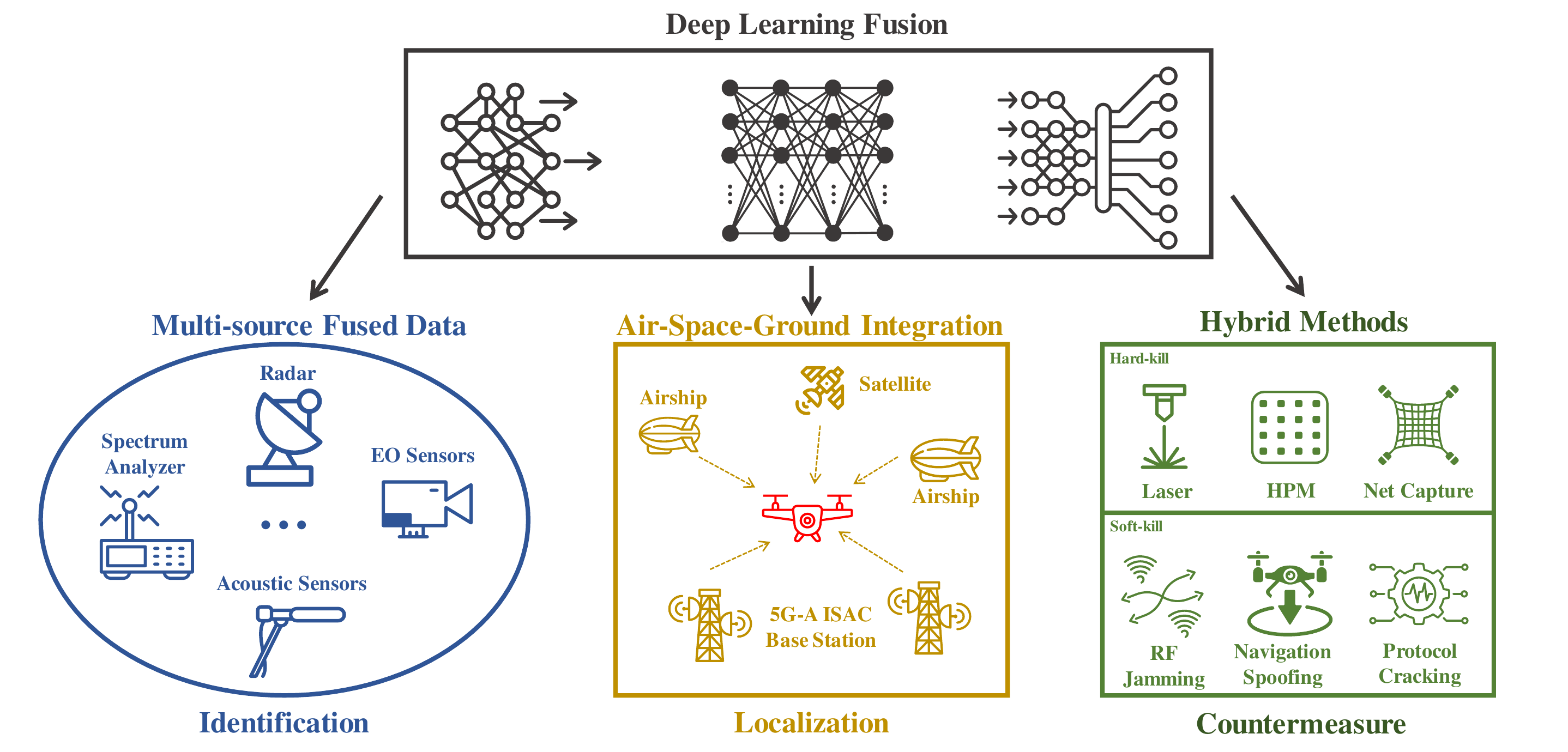}
		\caption{Deep learning-enhanced UAV management system.}
		\vspace{-4mm}
	\end{figure*}
	
	\section{Further Discussion}
	While numerous visions have been proposed for the future UAV management systems, there are still a series of challenges.
	The high-precision performance of deep learning models relies on massive volumes of labeled data. However, with the emergence of new UAVs, the problem of data scarcity has become increasingly prominent \cite{9779853}.
	More critically, UAVs are increasingly equipped with adversarial algorithms \cite{xu2025robust} such as dynamically encrypted communication signals and irregular flight trajectories to deliberately disrupt classification models.
	Furthermore, the weak signals emitted by low-altitude, slow-speed, and small-sized UAVs are difficult to capture.
	Confronted with these challenges, cutting-edge research directions are providing new impetus for the development of UAV management systems.
	
	\subsection{Federated Learning}
	As a distributed machine learning paradigm, federated learning enables collaborative model training across multiple data sources by exchanging model parameter updates without sharing raw data, which is crucial for addressing security concerns. Specifically, federated learning is deployed locally at each UAV management system node. Model updates from these nodes are aggregated on a central server after encryption, and the server generates a global model before redistributing it to all nodes, which enhances the model’s capability to identify novel UAV signals, especially in sensitive areas where data collection is challenging. However, non-independent and identically distributed data across nodes may affect the stability of model convergence, frequent model transmissions impose high demands on communication bandwidth, and strict encryption protocols must be designed to prevent privacy leakage \cite{11005535}.
	
	\subsection{Large Language Models}
	The natural language processing capabilities of large language models (LLMs) can be transferred to UAV management \cite{10643253}.
	After pre-training general-purpose LLMs and fine-tuning them with UAV domain knowledge, the vertical LLMs can leverage their internal knowledge for reasoning, enabling fast and accurate identification and behavior prediction. Dynamic adversarial training that simulates various communication encryption modes and abnormal flight trajectories can proactively enhance the model’s anti-interference capability.
	When encountering UAVs with entirely new encryption protocols or flight patterns, the vertical LLMs can leverage internal knowledge for reasoning, achieving fast and accurate identification and behavior prediction. Nevertheless, LLM training and inference consume computational resources, and the interpretability of decision-making processes requires further attention.
	
	\subsection{Quantum Receivers}
	Quantum receivers represent a disruptive direction in next-generation sensing technology, capable of breaking through the sensitivity limits of UAV detection \cite{10845209}.
	Unlike conventional sensors that are susceptible to interference in complex electromagnetic environments, quantum receivers leverage the high sensitivity of atoms to electromagnetic fields to detect extremely weak signals that are imperceptible to conventional systems. Combined with AI-powered adaptive filtering technology, they can effectively extract signals from strong background noise, improving the detection range and accuracy.
	However, current cutting-edge devices are costly and require extreme laboratory conditions such as ultra-low temperatures and high vacuum. Achieving miniaturization, high stability, and practical deployment is still challenging.
	
	\section{Conclusions}
	This paper has systematically presented an integrated UAV management system that synergistically combines multi-sensor data fusion identification, high-precision localization, and adaptive countermeasures, as shown in Figure 12.
	The core contribution lies in establishing a closed-loop intelligent UAV management framework, namely an integrated UAV identification, localization, and countermeasure design based on multimodal multi-sensor data.
	By leveraging deep learning techniques across these stages, the proposed UAV management system demonstrated significant improvements in identification accuracy, dynamic tracking robustness, and countermeasure efficiency.
	Looking forward, several critical challenges remain to be addressed to advance this field.
	Future research should focus on overcoming data scarcity for UAVs, enhancing model robustness against adversarial signals and dynamic flight patterns, and improving weak signal detection capabilities under complex electromagnetic conditions.
	Promising directions include incorporating federated learning for privacy-preserving collaboration, leveraging LLMs for few-shot generalization, and integrating quantum receivers to achieve breakthroughs in sensitivity and anti-jamming performance.

\end{document}